\documentstyle[prb,twocolumn,aps]{revtex} 

\input{psfig}
\begin{document}
\draft

\twocolumn[\hsize\textwidth\columnwidth\hsize\csname@twocolumnfalse\endcsname

\title{Local Density of States in a Dirty Normal Metal Connected to a 
Superconductor}

\author{W. Belzig$^1$, C. Bruder$^1$, and Gerd Sch\"on$^{1,2}$} 

\address{$^1$Institut f\"ur
  Theoretische Festk\"orperphysik, Universit\"at Karlsruhe, D-76128
  Karlsruhe, Germany} 
\address{$^2$Department of Technical Physics, Helsinki University of 
  Technology, FIN-02150 Espoo, Finland}

\maketitle
\date{\today}
\begin{abstract}
  A superconductor in contact with a normal metal not only induces
  superconducting correlations, known as proximity effect, but also
  modifies the density of states at some distance from the interface.
  These modifications can be resolved experimentally in
  microstructured systems.  We, therefore, study the local density of
  states $N(E,x)$ of a superconductor - normal metal heterostructure.
  We find a suppression of $N(E,x)$ at small energies, which persists
  to large distances.  If the normal metal forms a thin layer of
  thickness $L_n$, a minigap in the density of states appears which is
  of the order of the Thouless energy $\sim \hbar D/L_n^2$.  A
  magnetic field suppresses the features.  We find good agreement with
  recent experiments of Gu\'eron {\it et al.}
\end{abstract}

]
\section{Introduction}
A normal metal in contact with a superconductor acquires partially
superconducting properties.  Superconducting correlations, described
by a finite value of the pair amplitude $\langle
\psi_{\downarrow}(\bbox{x}) \psi_{\uparrow}(\bbox{x}) \rangle$,
penetrate some distance into the normal metal.  This {\it proximity
effect} has been studied since the advent of BCS theory (see 
Ref.~\onlinecite{deutscher} and references therein). Recently, progress in
low-temperature and microfabrication technology has rekindled the
interest in these properties \cite{mesosc,m2,m3,m4,pothier:96}.
Interference effects in a dirty normal metal increase the Andreev
conductance \cite{hekkingnazarov,pothier94a}. The effect of the
superconductor on the level statistics of a small normal grain has
been investigated \cite{beenakkermelson}.

Whereas the order parameter penetrates into the normal metal, the pair
potential $\Delta(\bbox{x})$ vanishes in the ideal metal without
attractive interaction.  Since $\Delta$ yields the gap in the
single-particle spectrum of a bulk superconductor, the question arises
how the spectrum of the normal metal is modified by the proximity to
the superconductor. Recently, this question has been investigated
experimentally by Gu\'eron {\it et al.} \cite{pothier:96}.  In their
experiment, the local density of states of a dirty normal metal in
contact with a superconductor was measured at different positions and
as a function of an applied magnetic field.

In this paper, we evaluate the local density of states $N(E,x)$ of a
superconductor - normal metal heterostructure with impurity scattering
in a variety of situations.  We generalize earlier theoretical work
\cite{mcmillan:68,golubov:88,golubov:95,dyachenko} by applying the
quasiclassical Green's function formalism and by including the effect
of a magnetic field. We compare with the experiment of Gu\'eron {\it
et al.}  \cite{pothier:96} and find good qualitative agreement with
the experimental data both in the cases with and without a magnetic
field.

\section{The model}
\begin{figure}
  \centerline{\psfig{figure=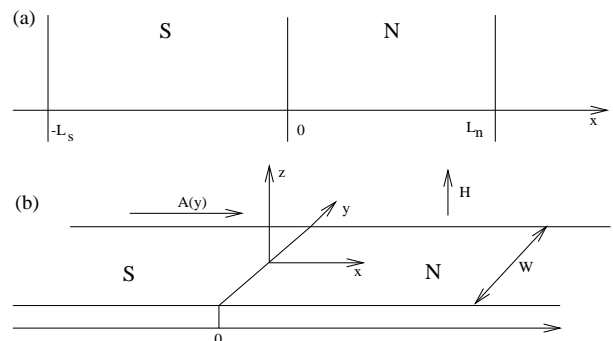,width=8cm}}
  \caption[]{\label{geometry}Geometries considered in this 
    article. (a) A strictly one-dimensional geometry. (b) A more realistic 
    geometry similar to experimental setup.}
\end{figure}

In the following we will consider geometries as shown in
Fig.~\ref{geometry}. The superconductor is characterized by a finite
pairing interaction $\lambda$ and transition temperature $T_c > 0$. In
the normal metal we take $\lambda=T_c=0$.  Here we restrict ourselves
to the dirty (diffusive) limit, $\xi \gg l_{\text{el}}$, where
$\xi=(D/2\Delta)^{1/2}$ is the superconducting coherence length at
$T=0$ and $l_{\text{el}}$ is the elastic mean free path.  The latter
is related to the diffusion constant via
$D=\frac{1}{3}v_Fl_{\text{el}}$.
 
The density of states (DOS) of this inhomogeneous system can be
derived systematically within the quasiclassical real-time Green's
functions formalism \cite{eliashberg}.  In the dirty limit the
equation of motion for the retarded Green's functions $G_E$, $F_E$ reads
\cite{usadel:70}
\begin{eqnarray}
  \label{usadel.real} 
  & &\frac{D}{2}\,\left(
    G_E\,(\vec{\nabla}-2ie\vec{A})^2 F_E-
    F_E\,\vec{\nabla}^2 G_E\right)=\\
  \nonumber & &
  (-iE+\Gamma_{\text{in}})\,F_E-\Delta\,G_E+2\Gamma_{\text{sf}}\,G_E\,F_E.
\end{eqnarray}
The diagonal and off-diagonal parts of the matrix Green's function,
 $G_E$ and  $F_E$, obey the normalization condition
\begin{equation}
  \label{normal}
  G_E^2+F_E^2=1\;,
\end{equation}
which suggests to parameterize them by a function $\theta(E,x)$ via
$F_E=\sin(\theta)$ and $G_E=\cos(\theta)$.
Inelastic scattering processes are accounted for by the rate
$\Gamma_{\text{in}}=1/2\tau_{\text{in}}$, while scattering processes from
paramagnetic impurities are described by the spin-flip rate 
$\Gamma_{\text{sf}}=1/2\tau_{\text{sf}}$. At low temperatures 
the former is very small ($\Gamma_{\text{in}}\sim 10^{-3}\Delta$), 
and will be neglected in the following.

For the geometry shown in Fig.~\ref{geometry} the order parameter can
be taken real. On the other hand, in the vicinity of an N-S boundary
the absolute value of the order parameter is space dependent, and has
to be determined self-consistently.  The self-consistency condition is
conveniently expressed in the imaginary-time formulation, where
\begin{equation}
  \label{selfcons.general}
  \Delta(x)\ln (\frac{T}{T_c(x)})=
  2\pi T\sum_{\omega_\mu>0}
  F_{i\omega_\mu}(x)-\frac{\Delta(x)}{\omega_\mu}\; .
\end{equation}
Here, $\omega_\mu = \pi T(2\mu+1)$ are Matsubara frequencies.  The
summation is cut off at energies of the order of the Debye energy.
The coupling constant in S has been eliminated in favor of $T_c$,
while the coupling constant in N is taken to be zero.

In the case where the interface between N and S has no additional
potential, the boundary conditions are
\cite{kuprianov:88}
\begin{eqnarray}
  \label{bound.cond}
  F_E(0_-) & = &  F_E(0_+) \\ \nonumber
  \frac{\sigma_s}{G_E(0_-)}\frac{d}{dx}F_E(0_-) & =&
  \frac{\sigma_n}{G_E(0_+)}\frac{d}{dx}F_E(0_+)\; .
\end{eqnarray}
Here, $\sigma_{n(s)}$ are the conductivities of the normal metal and
the superconductor, respectively.  The complete self-consistent
problem requires a numerical solution.  Starting from a step-like
model for the order parameter, self-consistency was typically reached
within 10 steps.  Finally the DOS is obtained from
$ N(E)=N_0 Re G_E(x)$,
where $N_0$ is the Fermi level DOS in the normal state.

We will present now results for three different cases:
\begin{itemize}
\item[A.] The DOS near the boundary of a
  semi-infinite normal metal and superconductor.
\item[B.] The DOS in a thin normal film in contact with a bulk superconductor.
\item[C.] The effect of a magnetic field on the DOS in an
experimentally realized N-S heterostructure.
\end{itemize}
In the following sections energies and scattering rates will be measured in 
units of the bulk energy gap $\Delta$ and distances in units of the 
coherence length $\xi=(D/2\Delta)^{1/2}$.

\section{Results and discussion}
\subsection{DOS in an Infinite System}
We assume that the normal metal and the superconductor are much
thicker than the coherence length $L_s,L_n\gg\xi$ and investigate how
the DOS changes continuously from the BCS form
$N_{\text{BCS}}(E)/N_0=\left|E\right|/(E^2-\Delta^2)^{1/2}$ deep
inside the superconductor to the constant value $N_N(E)/N_0=1$ in the
normal metal.

In a first approximation, neglecting self-consistency and  
paramagnetic impurities, we can solve Eq.~\ref{usadel.real}
analytically, with the result
\begin{equation}
  \label{usadel.analytic}
  \theta(E,x)=\left\{
      \begin{array}{l}
        4\text{atan}[\;\tan(\theta_0/4)
        \exp(-\sqrt{2\omega/D_n}\;x)\;] \quad\hfill x>0 
        \\[3mm]
        \theta_s+4\;\text{atan}[\;\tan((\theta_0-\theta_s)/4)
          \times
        \\ 
        \qquad
        \exp(\sqrt{2\sqrt{\omega^2+\Delta^2}/D_s}\;x)\;]
        \hfill  \quad x<0\;.
      \end{array}\right.
\end{equation}
Here
\begin{eqnarray*}
  \omega & = &-iE+\Gamma_{\text{in}},\\
  \theta_s & = & \text{atan}(\frac{\Delta}{-iE+\Gamma_{\text{in}}}),\\
  \sin\frac{\theta_0-\theta_s}{2} & =&
  \gamma\frac{(-iE+\Gamma_{\text{in}})^{1/2}}
  {((-iE+\Gamma_{\text{in}})^2+\Delta^2)^{1/4}}
  \sin\frac{\theta_0}{2}\;.
\end{eqnarray*}
Several  material parameters combine into the parameter 
\begin{equation}
\gamma=(\sigma_n\xi_s/\sigma_s\xi_n)\; ,
\end{equation}
measuring the mismatch in the conductivities and the coherence lengths
of the two materials. Furthermore, $\xi_{s(n)}$ is defined by
$(D_{s(n)}/2\Delta)^{1/2}$, where $D_{s(n)}$ is the diffusion constant
of the superconductor (normal metal).

\begin{figure}
  \centerline{\psfig{figure=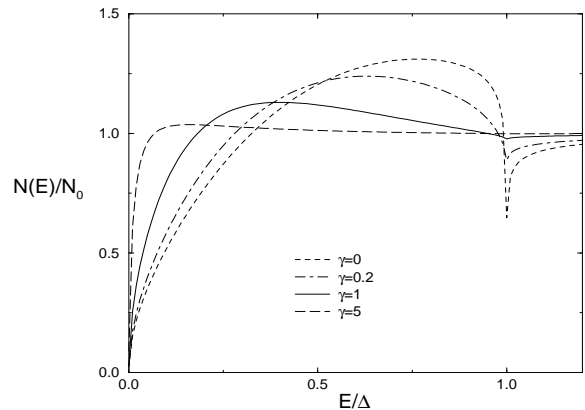,height=5.5cm,width=7cm}}
        \caption[]{\label{dos.x=2.gamma.eps}
          DOS in the normal metal at $x=1.5\xi_n$.}
\end{figure}

The resulting DOS, $N(E)$, in the normal metal at a distance
$x=1.5\xi_n$ from the interface is shown in
Fig.~\ref{dos.x=2.gamma.eps} for different values of the parameter
$\gamma$. It shows a sub-gap structure with a peak below the
superconducting gap energy $E < \Delta$ and a strong suppression at
zero energy.  The modification of the DOS is most pronounced at small
values of $\gamma$ and at small distances. The smaller the energy, the
larger is the distance where the modifications are still visible.  In
particular at $E=0$ the DOS vanishes for all values of $x$.
Pair-breaking effects lead to a finite zero-energy DOS, as will be
shown later.

\begin{figure}
  \centerline{\psfig{figure=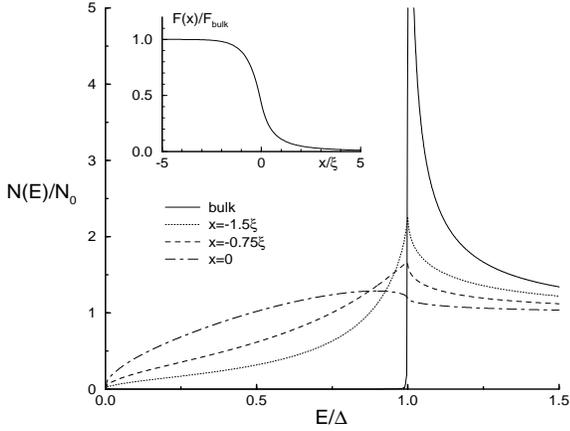,height=5.5cm,width=7cm}}
        \caption[]{\label{dos.inf.s}Density of states on the
          superconducting side of the N-S boundary. The inset shows 
          the self-consistent pair amplitude.}
\end{figure}

Next we solve the problem self-consistently and present some numerical
results for the case $\gamma=1$. We first concentrate on the
superconducting side of the boundary.  As shown in
Fig.~\ref{dos.inf.s} the peak in the DOS is strongly suppressed,
changing from a singularity to a cusp, but it remains at the same
position $\Delta$ as one approaches the boundary. On the other hand,
the density of states with energies below $\Delta$ increases.  The
states with energies well below $\Delta$ decay over a characteristic
length scale $\sqrt{D_s/(2\sqrt{\Delta^2-E^2})}$,
see Eq. (\ref{usadel.analytic}).
 
\begin{figure}
  \centerline{\psfig{figure=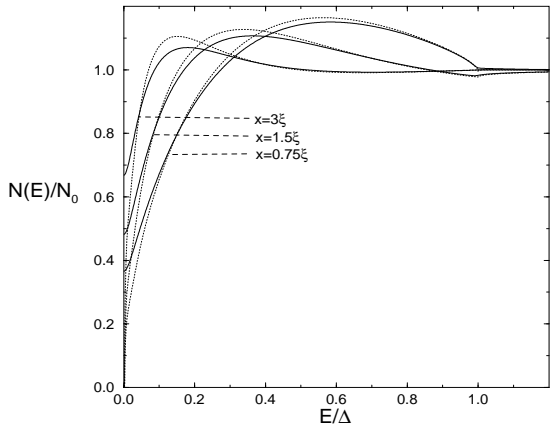,height=5.5cm,width=7cm}}
        \caption[]{\label{dos.inf.n}Density of states on the
          normal side of an N-S boundary for two spin-flip 
          scattering rates: $\Gamma_{\text{sf}}=0$ (dotted lines) and 
          $\Gamma_{\text{sf}}=0.015\Delta$ (solid lines).}
\end{figure}

The DOS on the normal side at different distances from the NS-boundary
is shown in Fig.~\ref{dos.inf.n}.  The pronounced sub-gap structure
found in the approximate solution is still present in the
self-consistent treatment.  The figure shows how the peak height and
position change with the distance.  In the absence of pair-breaking
effects the DOS vanishes at the Fermi level for all distances (dotted
curves).  Inclusion of a pair-breaking mechanism (solid curves)
regularizes the DOS at the Fermi level, and also the peak height is
suppressed.  The curves are in qualitative agreement with experimental
data shown in Ref.~\onlinecite{pothier:96}.  The self-consistent
calculation presented here leads to a slightly better fit than the
theoretical curves shown in Ref. \onlinecite{pothier:96} where a
constant pair potential was used in the solution of the Usadel
equation. In particular, the low-energy behavior of the experimental curves
is reproduced correctly.

At finite temperatures (but $T\ll T_c$) we expect no qualitative
changes in the behavior described above except that the structures in
the DOS will be smeared out by inelastic scattering processes. Hence
for an experimental verification temperatures as low as possible would
be most favorable.

\subsection{DOS in thin N-layers}
\begin{figure}
  \centerline{\psfig{figure=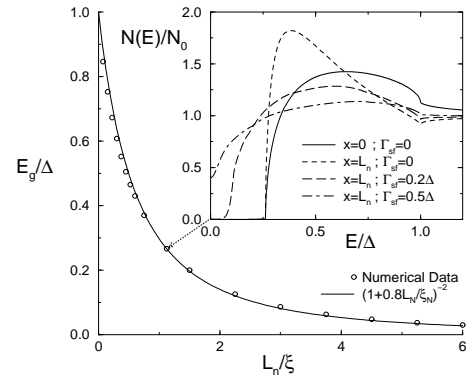,width=7cm}}
  \caption[]{\label{minigap}Minigap $E_g$ as a function of the 
    normal-layer thickness.  Inset: local DOS of an N-layer of
    thickness $L_n=1.1\xi$ in proximity with an bulk superconductor.}
\end{figure}

Next we consider a thin normal layer in contact with a bulk
superconductor, $L_s\gg L_n\simeq\xi$.  The boundary condition at 
$x=L_n$ is chosen to be $d\theta(E,x)/dx = 0$, i.e., the normal metal
is bounded by an insulator. In this case the DOS on the N
side develops a minigap at the Fermi energy, which is smaller than the
superconducting gap.  If the thickness of the normal layer is
increased, the size of this minigap decreases.  Results obtained from
the self-consistent treatment are shown in Fig.~\ref{minigap}.
Details of the shape of the DOS depend on the location in the N-layer
\cite{belzig}.  However, the magnitude of the minigap is
space-independent, as shown in the inset of Fig.~\ref{minigap}.  The
magnitude of the gap is expected to be related to the Thouless energy
$D/L_n^2$, which is the only relevant quantity which has the correct
dimension. Of course the relation has to be modified in the limit
$L_n\to 0$. Indeed as shown in Fig.~\ref{minigap} a relation of the
form $E_g \sim (\text{const}\  \xi +L_n)^{-2}$ fits quite well. The sum
of the lengths may be interpreted as an effective thickness of the
N-layer since the quasiparticle states penetrate into the
superconductor to distances of the order of $\xi$. 
The effect of spin-flip scattering in the normal metal on the minigap 
structure is also shown in the inset of Fig.~\ref{minigap}.
The minigap is suppressed as $\Gamma_{\text{sf}}$ is increased until 
a gapless situation is reached at $\Gamma_{\text{sf}}\approx 0.4 \Delta$.
We would like to mention that a similar feature had been found before
by McMillan \cite{mcmillan:68} within a tunneling model ignoring the
spatial dependence of the pair amplitude.  We have considered here the
opposite limit, assuming perfect transparency of the interface but
accounting for the spatial dependence of the Green's functions.  
For $\Gamma_{\text{sf}}=0$ our results for the structure of the DOS agree 
further with previous findings of Golubov and Kupriyanov \cite{golubov:88} 
and Golubov {\it et al.} \cite{golubov:95}.  Recently, a minigap in a 
two-dimensional electron gas in contact to a superconductor has also been
studied\cite{volkovetal94}.

\subsection{Density of states in a magnetic field}
An applied magnetic field suppresses the superconductivity in both
superconductor and normal metal. To study the effect of the magnetic
field on our system we consider the geometry shown in
Fig.~\ref{geometry}b.  Because in the experimental setup the
thickness of the films is much smaller than the London penetration
depth, we can neglect the magnetic field produced by screening
currents. Therefore it is reasonable to assume a constant magnetic
field, which is present in both S and N.  The vector potential is
then chosen to be
\begin{equation}
  \label{vecpotB}
  \vec{A}=A(y)\vec{e}_x\quad;\quad
  A(y)=Hy \; .
\end{equation}
Eq.~(\ref{usadel.real}) can be considerably simplified in the case that
the size of the system in $y$-direction is smaller or of the order of
$\xi$. The system is limited to $-W/2<y<W/2$, where $ W \simeq \xi$.
Therefore the Green's functions do not depend on $y$ and the equation
can be averaged over the width $W$. The equation reduces to the
effective one-dimensional equation
\begin{eqnarray}
  \label{usadel.eff}
  & &
  \frac{D}{2}\,\left(G_E\,\partial_x^2 F_E-F_E\,\partial_x^2 G_E\right)=
  \\ \nonumber
  & & (-iE+\Gamma_{\text{in}})\,F_E-\Delta\,G_E+
2\Gamma_{\text{eff}}\,G_E\,F_E\;.
\end{eqnarray}
Here, $\Gamma_{\text{eff}}=\Gamma_{\text{sf}}+De^2H^2W^2/12$
acts as an effective pair-breaking rate, which depends on the
transverse dimension and the applied magnetic field.

\begin{figure}
  \centerline{\psfig{figure=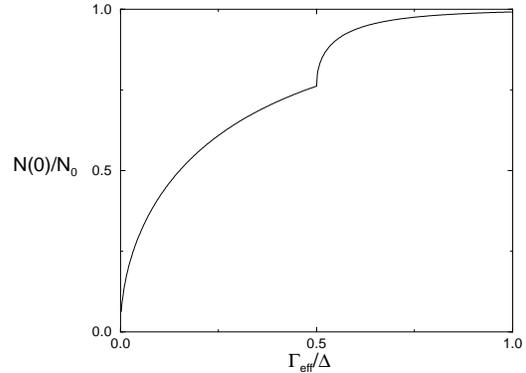,width=7cm}}
  \caption[]{\label{dos0}
    Zero-energy DOS in the normal metal at $x=1.5\xi$ 
    as a function of $\Gamma_{\text{eff}}$.}
\end{figure}

If we approximate the Green's functions in the superconductor by their bulk 
values, the DOS in the normal metal at zero energy can be 
calculated analytically:
\begin{equation}
  \label{zero.energy.dos}
  \frac{N(0)}{N_0}=\left\{
      \begin{array}{lr}
        \text{tanh}(2\sqrt{\Gamma_{\text{eff}}/D}\;x) &
        2\Gamma_{\text{eff}}<\Delta\\[3mm]
        (1-\alpha^2)/(1+\alpha^2)
        & 2\Gamma_{\text{eff}}>\Delta 
      \end{array}\right.,
\end{equation}
where
\begin{equation}
  \label{alpha}
  \alpha=\frac{\Delta \exp(-2\sqrt{\Gamma_{\text{eff}}/D}\;x)}
  {2\Gamma_{\text{eff}}+\sqrt{4\Gamma_{\text{eff}}^2-\Delta^2}}\;.
\end{equation}
In Fig.~\ref{dos0} the dependence of the DOS on $\Gamma_{\text{eff}}$
at $x=1.5\xi$ is shown for two different spin-flip scattering rates
(equal rates for normal metal and superconductor).  At
$\Gamma_{\text{eff}}=0.5\Delta$ the field dependence of the DOS shows
a kink.  This kink arises because above this value of
$\Gamma_{\text{eff}}$ the zero-energy DOS in the superconductor is
nonzero (gapless behavior), which leads to an even stronger
suppression of the proximity effect.

\begin{figure}
  \centerline{\psfig{figure=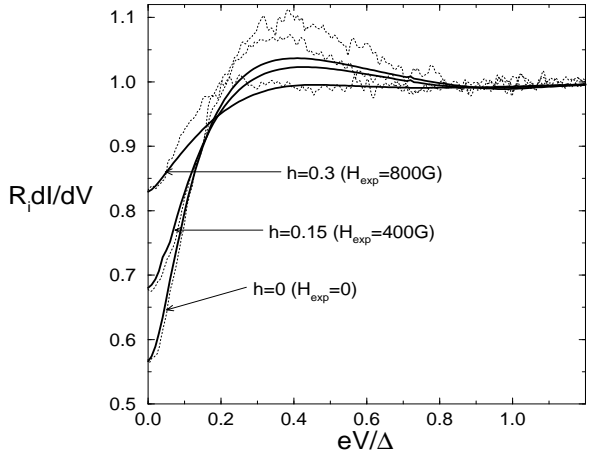,width=9cm,height=7cm}}
  \caption[]{\label{dos.gamma.eff} Quantitative comparison of
  ex\-per\-i\-ment\protect\cite{pothierunpub} (dotted lines) and
  theory (solid lines). The experimental magnetic fields are $H=0$, 
  $400$, and $800$G; $h=HeW(D/12\Delta)^{1/2}$. The theoretical curves have
  been normalized by $R_i\equiv dI/dV_{\text{exp}}(eV/\Delta=1.5)$.}
\end{figure}

Figure~\ref{dos.gamma.eff} shows a quantitative comparison of these
results with experimental data taken by the Saclay
group\cite{pothierunpub}.  In this experiment \cite{pothier:96}, the
differential conductance of three tunnel junctions attached to the
normal metal part of the system was used to probe the DOS at different
distances from the superconductor.  Accordingly, we have calculated
the self-consistent DOS in the presence of a magnetic field throughout the
system for all energies and determined the
differential conductance \cite{p(e)}.  We used $x=1.8\xi$, consistent
with an estimate from a SEM-photograph, and used a spin-flip
scattering rate of $\Gamma_{\text{sf}}=0.015\Delta$ in the normal
metal as a fit parameter. This is necessary in the framework of our
approach to explain the finite zero-bias conductance at zero field.
We, furthermore, assumed ideal boundary conditions at the NS interface,
i.e., $\gamma=1$, the motivation being that great care was used in the 
experiment to produce a good metallic junction, and significant Fermi 
velocity mismatches are not to be expected.  

At low and high voltages the agreement with the experimental data is good
for all three field values. On the other hand, the maximum in 
the DOS is not reproduced well by our calculation.  Including the effect of a 
non-ideal boundary, i.e., $\gamma<1$ leads to an increase of the peak in the 
DOS but to a less satisfactory fit at low voltages. 
We cannot resolve this discrepancy, but we would like to point out that
our theory is comparatively simple and does not include all the
geometric details of the experiment  (e.g., the geometry of the overlap 
junction is not really one-dimensional and would be difficult to treat 
realistically). Our intention is to show that theoretical treatment described
here contains the physical ingredients to explain the basic features of the 
experimental data. The overall agreement between theory and experiment
demonstrated in Fig. 7 shows this to be the case.

\section{Conclusions and Outlook}
In conclusion, we have given a theoretical answer to the question
asked in the introduction, viz., what is -- beyond the proximity
effect -- the effect of a superconductor on the spectrum of a normal
metal coupled to it. Using the (real-time) Usadel equations, we have
calculated the local density of states in the vicinity of an N-S
boundary in both finite and infinite geometries.  It shows an
interesting sub-gap structure: if the normal metal is infinite, the
density of states is suppressed close to the Fermi energy, but there
is no gap in the spectrum. This is the behavior found in a recent
experiment \cite{pothier:96}. In thin normal metals we find a mini-gap
in the density of states which is of the order of the Thouless energy.
We have also investigated the suppression of these effects by an
applied magnetic field and find good agreement with experiment.

We are grateful to D. Esteve and H. Pothier for raising the
questions leading to this work and for many inspiring discussions. We
would also like to acknowledge helpful discussions with N.~O.
Birge, M. Devoret, S. Gu\'eron, and A.~D. Zaikin. The support of the 
Deutsche Forschungsgemeinschaft, through SFB 195, as well as the 
A.~v.~Humboldt award of the Academy of Finland (GS) is gratefully acknowledged.


\begin{references}
\bibitem{p(e)} The differential conductance also shows single-electron
effects caused by the small capacity of the tunnel junctions. It is
necessary to include them for a quantitative comparison, and we have
done so using the prescription given in Ref.~\protect\onlinecite{pothier:96}.
\end{references}
\end{document}